**Strong and broadly tunable plasmon resonances in thick films of aligned carbon nanotubes**

Kuan-Chang Chiu[1,2], Abram L. Falk[1,*], Po-Hsun Ho[1], Damon B. Farmer[1], George Tulevski[1], Yi-Hsien Lee[2], Phaedon Avouris[1], Shu-Jen Han[1]

[1]IBM T. J. Watson Research Center, Yorktown Heights, New York, 10598, USA

[2]Dept. of Mat. Sci. and Eng., National Tsing-Hua University, Hsinchu, 30013 Taiwan

**Abstract:**

Low-dimensional plasmonic materials can function as high quality terahertz and infrared antennas at deep subwavelength scales. Despite these antennas' strong coupling to electromagnetic fields, there is a pressing need to further strengthen their absorption. We address this problem by fabricating thick films of aligned, uniformly sized carbon nanotubes and showing that their plasmon resonances are strong, narrow, and broadly tunable. With thicknesses ranging from 25 to 250 nm, our films exhibit peak attenuation reaching 70%, quality factors reaching 9, and electrostatically tunable peak frequencies by a factor of 2.3×. Excellent nanotube alignment leads to the attenuation being 99% linearly polarized along the nanotube axis. Increasing the film thickness blueshifts the plasmon resonators down to peak wavelengths as low as 1.4 μm, promoting them to a new near-infrared regime in which they can both overlap the $S_{11}$ nanotube exciton energy and access the technologically important infrared telecom band.

**Keywords:** Carbon nanotube, plasmon, resonator, nanophotonics, infrared, multispectral

**Main text:**

The Fabry-Pérot plasmon resonances of carbon nanotubes, longitudinal charge oscillations bound by the nanotube ends[1–6], have remarkable properties and a strong technological potential. They can concentrate light into nanoscale volumes[3] and have a natural application to surface-enhanced infrared absorption (SEIRA) spectroscopy[6,7]. Moreover, along with analogous resonances in graphene[8–13], they are a promising basis for a new class of fast and efficient photothermoelectric photodetectors[14–16]. Tunability with length[17,18] and free-charge density[19,20] has allowed nanotube resonances to span wavelengths from the terahertz/far-infrared[17–21] (as high as 200 μm[21]) to the mid-infrared[6], to even the near-infrared (down to 1.4 μm), as we show in this work. Taken together, these properties make nanotube plasmonics a promising foundation for multispectral cameras that use a single photosensitive material to achieve vision through the entire infrared range.

---

[*] alfalk@us.ibm.com

However, in order for photodetectors comprising low-dimensional materials to be externally efficient, a key hurdle is achieving high absorption. Strategies to enhance absorption have included incorporating graphene into external antennas[22] or optical cavities[23], engineering interference with substrate dielectrics[24], and using multilayer graphene or graphene/insulator stacks[25]. Each has its own advantages and drawbacks. For instance, on-chip interference could lead to total light absorption[24] but constrains the resonator shape and substrate type. Solution-processed thick films of carbon nanotubes offer a particularly clear-cut route to high absorption. In addition, semiconducting nanotubes can be purified in solution and exhibit a higher photothermoelectric coefficient than either metallic nanotubes or graphene[26].

For this Letter, we assembled thick films of uniform nanotube-plasmon resonators, resulting in strong absorption and narrow ensemble linewidths. The peak attenuation that we observe (up to 70%) is markedly higher than both the ~2% peak attenuation observed in a thin film ($t = 6$ nm) of nanotube resonators[6] and the ~6% peak attenuation typically seen in graphene nanoribbons[9]. The $Q$ factors are as high as 9, attenuation is 99% linearly polarized, and the plasmon-resonance frequencies are electrostatically tunable by a factor of 2.3×, a higher factor than the 1.4× observed[6] in thin nanotube films. We tune the plasmon resonators through the nanotube's $S_{11}$ exciton to wavelengths as low as 1.4 μm, half the wavelength of previously fabricated nanotube resonators. Our results show that nanotube-plasmon resonators provide an exciting pathway toward efficient and broadband infrared cameras, compact chemical sensors, and optoelectronics at deep subwavelength scales.

To fabricate these films, we made use of a controlled vacuum filtration method[27] to produce thick films of remarkably well-aligned nanotubes (Fig 1(a) and Supporting Information (SI)). We dispersed semiconducting nanotubes in an aqueous solution with the surfactant sodium dodecylbenesulfonate (SDBS), sonicated and centrifuged the solution, and filtered the supernatant through polycarbonate filter membranes at a very low filtration speed (1 ml/hr). The nanotube alignment is templated by grooves in the filter paper, with the overall degree of alignment determined by the competition between van der Waals forces, pressure from vacuum pumping, and electrostatic interactions between nanotubes deriving from their surface charge.

This method is simple, low-cost, and scalable up to large areas. Over our 1-inch filters, it reliably produced uniform films of globally aligned nanotubes, with the thickness ($t$) ranging from a moderate 25 nm to an ultrathick 250 nm.

To unify the nanotube length, we transferred the films to high-resistivity silicon substrates (Fig. 1(b)-(c)), dissolved the filter paper that had supported them, and used electron-beam lithography and reactive ion oxygen etching to pattern the films into stripes of uniform nanotube segments, with segment length ($L$) ranging from 100 nm to 800 nm (Fig. 1(d)-(g)). In atmospheric conditions, water absorption naturally induces positive charge carriers in the nanotubes[28]. We also exposed the nanotubes to either $NO_2$ gas or $HNO_3$ vapor, strong oxidizers that induce much higher free-charge densities[29]. We used micro Fourier-transform infrared spectroscopy (μ-FTIR) to measure the films' attenuation at room temperature, with each measurement incorporating 10-100 million nanotubes in a ~(50 μm)$^2$ area. For films with resonant frequencies exceeding 6000 cm$^{-1}$, we measured attenuation in a (5 mm)$^2$ area, using an ultraviolet-visible-near-infrared (UV-VIS-NIR) spectrometer.

We observe prominent, narrow attenuation peaks ($\nu_p$) centered at mid- and near-infrared frequencies (Fig. 2). Three key characteristics prove that these peaks indeed correspond to nanotube-plasmon resonance. First, the center frequency of the peak is a strongly decreasing function of $L$ (Fig. 2(a)), as expected from a Fabry-Pérot resonance. Second, exposure to $NO_2$ gas causes the resonance to blue-shift and to intensify (Fig. 2(b)), as expected from a plasmon resonator whose free charge density is increasing. As the $NO_2$ desorbs over a period of several days, the plasmon attenuation shifts back to its original curve. Finally, we can separately observe both the plasmon resonance and the $S_{11}$ exciton peak (Fig. 2(c) and 3), verifying that the plasmon resonance peak is distinct from $S_{11}$. These characteristics confirm that the $\nu_p$ attenuation peaks correspond to plasmon resonance.

Thick-film-nanotube plasmon resonators are notably blueshifted from those in thin films[6] with the same $L$. This shift can be understood as a coupled antenna effect[30,31], in which charges in a given nanotube are accelerated by the electric field deriving from plasmons in neighboring nanotubes. In one of our thickest films ($t$ = 220 nm, $L$ = 200 nm), the plasmon resonance frequency reaches 7000 cm$^{-1}$ (Fig. 3), over twice the 3000 cm$^{-1}$ frequency observed[6] in a highly doped thin film of much shorter nanotubes ($t$ = 6 nm and $L$ = 30 nm), and a factor of 2.3× the resonance frequency of the same film when undoped (Fig. 3, inset). Altogether, the $\nu_p$ frequencies that we observed ranged from 1000 cm$^{-1}$ (Fig. S2 in SI) up to 7000 cm$^{-1}$, or equivalently, 1.4 to 10 µm. By reaching both the $S_{11}$ exciton energy and the 1.55-µm telecom wavelength, nanotube plasmon resonators can now be applied to make nanophotonic and optoelectronic devices in this technologically important range.

Because nearly all applications of plasmonics are limited by dissipation, an important metric is the $Q$ factor, which is inversely proportional to the ensemble dissipation rate. Calculating $Q = \nu_p/\Delta\nu_p$, where $\Delta\nu_p$ is the full-width-at-half-maximum of the plasmon resonance centered at $\nu_p$, we find our films to have $Q$ factors as high as 9.0 (inset to Fig. 2(a)). The trend of decreasing $Q$ factor with increasing $L$ can be understood to derive from the decrease in nanotube-length uniformity as $L$ approaches the average nanotube length in our solution, which is ~500 µm. Our $Q$ factors are significantly higher than those previously observed in thin nanotube films[6] ($Q \sim 3$ when not hybridized with phonons), and even higher than the $Q \sim 5$ typically seen in graphene nanoribbons[9]. These high $Q$ factors reflect the excellent nanotube alignment that vacuum filtration produces.

Peak attenuation, another metric of our film's performance, is as high as 70% (Fig. 2(a) and Fig. S3 in SI) of the polarized incident light. Due to limitations of our lithography and etching processes, our films consist of nanotubes only covering 30-70% of the chip surface in the measurement area (see. Fig. 1(g)). This limitation is responsible for the decreasing peak attenuation with decreasing $L$ seen in Fig. 2(a). If we were to normalize the attenuation by a geometrical factor accounting for this partial coverage, the normalized attenuation would reach >95%. In the future, simply improving the verticality of the etching process ought to allow nearly total attenuation in thick nanotube films.

For films with lower frequency resonances (i.e. approaching 1000 cm$^{-1}$), the lineshapes are asymmetrical (Fig. 4). We fit these curves to Breit-Wigner-Fano lineshapes[32,33]:

$$A(\nu) = I \frac{(F\gamma + \nu - \nu_0)^2}{(\nu - \nu_0)^2 + \gamma^2} \quad (1)$$

with *I* the intensity, $\nu_0$ and $\gamma$ the position and linewidth of the resonance, respectively, and *F* the Fano parameter, which characterizes the resonance's degree of asymmetry. The $A(\nu)$ fits show excellent agreement to the experimentally measured lineshapes. These Fano lineshapes can be understood as interference between the plasmons and LO phonons in the substrate[6] (see SI for further discussion). More generally, this type of plasmon-phonon coupling indicates that thick nanotube films could be used for surface-enhanced infrared absorption (SEIRA)[34–36], in which the nanotube plasmons strengthen the vibrational absorption from nearby molecules.

Excellent alignment of our nanotubes leads to the plasmon attenuation being nearly entirely polarized. To precisely determine the degree of polarization, we fit $A(\nu)$ lineshapes to Eq. 1 as a function of incident light polarization ($\theta$) (Fig. 4). The overall degree to which the attenuation is linearly polarized, $(I_\parallel - I_\perp) / (I_\parallel + I_\perp)$, with $I_\parallel$ ($I_\perp$) denoting the fitted intensity parallel (perpendicular) to the axis of the nanotubes, is 99.3%. Thick nanotube films are therefore an ideal material system for infrared and terahertz polarizers[37], as well as metamaterials requiring an extremely asymmetric dielectric constant[38].

For low-dimensional materials, the general shape of the plasmon-dispersion curve is determined by geometrical considerations. As a 1D system, isolated nanotubes are expected to have an approximately linear $\nu$ vs $q$ relationship[2], where the wavevector $q = \pi / L$. For graphene nanoribbons, a 2D system, $\nu \propto \sqrt{q}$.[9] For our thick films, $\nu$ exhibits a notably sublinear dependence with both $q$ (Fig. 5(a)) and $t$ (Fig. 5(b),(c)). Fitting both *L*- and *t*-dependence together, we find that $\nu$ is well-fit to the function:

$$\nu_p = \nu_0 \sqrt{tq}, \qquad (2)$$

with $\nu_0$ = 3335 cm$^{-1}$ immediately after NO$_2$ exposure (Fig. 5(b),(c)). Over time, as the films' charge density decreases, $\nu_0$ decreases by a factor of ~2. While Eq. 2 is undoubtedly an approximation, it clearly indicates that in thick films, inter-nanotube coupling causes the shape of the dispersion curve to deviate from that of an isolated 1D plasmonic system. From a technological perspective, Eq. (2) provides a means by which the resonant frequency of nanotube films can be rationally controlled.

Different types of nanotube films may find different applications. Thick and/or dense nanotube films are ideal for near-infrared detectors, since their resonant frequencies are higher, whereas thin or loosely packed nanotube films would allow analytes to intercalate into the film and be better for SEIRA. Thin films are more amenable than thick films to being tuned with an electrical gate, though interestingly, the 2.3× increase in $\nu_p$ that we observe with charge-transfer doping in our very thick ($t = 220$ nm) film is significantly higher than the factor of 1.4× observed in $t = 6$ nm films[6] and $t = 40$ nm films (Fig. S2 in SI). An important open question is the role of plasmon tunneling in dense nanotube films: are the plasmons' charges confined to individual nanotubes, or do they tunnel across nanotube-nanotube interfaces? A related phenomenon, the hot-carrier photothermoelectric effect[16], is a promising mechanism for high-speed photodetectors in which excited charges tunnel into metal contacts.

In conclusion, we showed that thick films of uniformly sized carbon nanotubes can support coherent plasmon resonances that are highly absorptive, highly polarized, and tunable up to near-infrared telecom frequencies. The key to developing these resonators into high-performance infrared- and terahertz-frequency photodetectors will now be engineering the interface from nanotubes to metal contacts. One promising approach would be to anneal refractory metal contacts[39] that form an end-bonded carbide to the thick nanotube film. Nanotube-plasmon resonators could also create quasi-coherent light sources when they concentrate thermal emission or electroluminescence into their resonant frequency[40], or even plasmonic lasers[13]. Together, these detectors and light sources would make carbon nanotubes a pathway to integrated multispectral nanophotonics.

**Acknowledgments:** The authors thank Javier García de Abajo and Jerry Tersoff for discussions, and Jim Bucchignano for performing the electron-beam lithography.

**Supporting Information:** Detailed procedure for fabricating thick films of aligned carbon nanotubes, spectroscopy methods, supporting discussion of background subtraction and $Q$ factor calculation, and supporting figures portraying plasmon attenuation in thinner nanotube films.

**Conflict of interest disclosure:** The authors declare no competing financial interest.

**Figures**

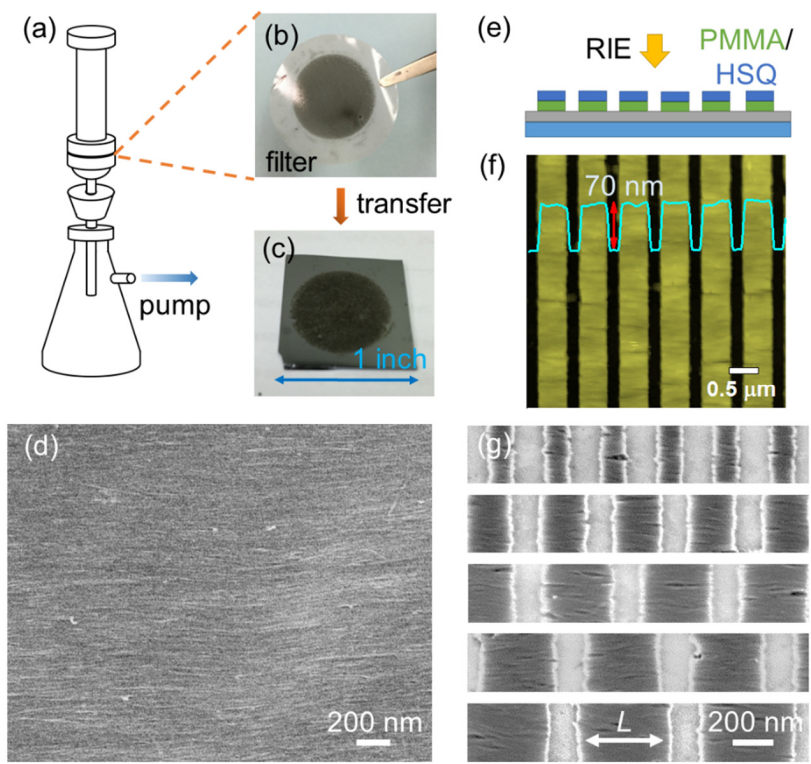

**Figure 1.** Preparation of thick films of aligned, uniformly sized carbon nanotubes. (a) Slow vacuum filtration produces (b) thick films of aligned carbon nanotubes on filter paper. (c) The nanotube films are transferred to silicon and exhibit (d) global alignment across the 1" circle in scanning electron microscopy (SEM). (e) The nanotubes are cut into stripes by reactive ion etching (RIE) through a poly(methyl methacrylate) / hydrogen silsesquioxane (PMMA/HSQ) mask, which was patterned with electron beam lithography. (f) Atomic force microscopy shows film thickness ($t$) to range from 25 to 250 nm. (g) SEM image of cut, aligned thick films of nanotubes.

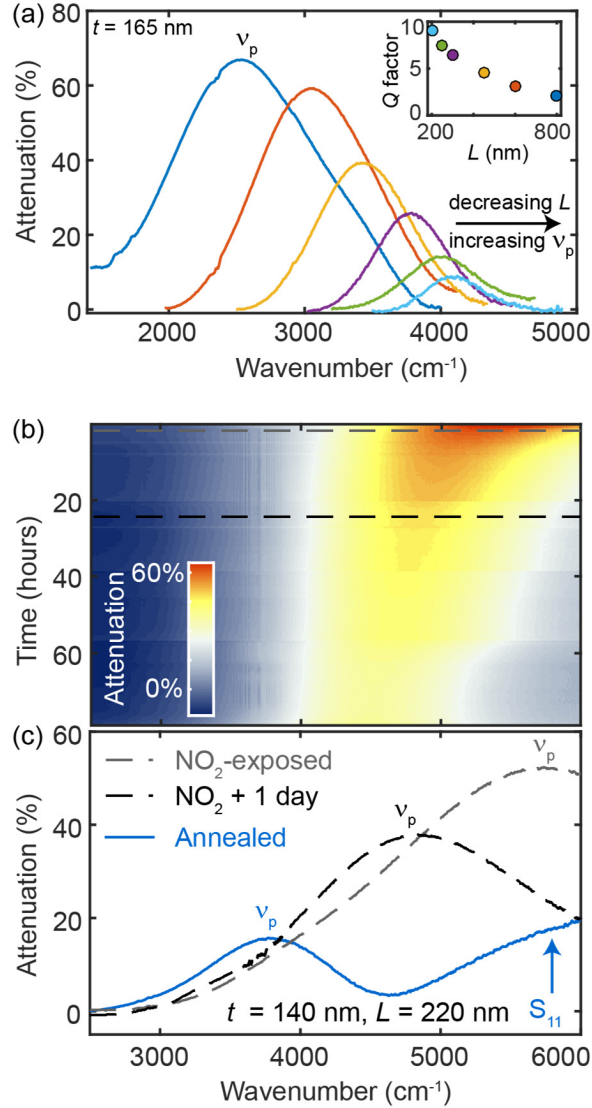

**Figure 2.** Attenuation from thick-film nanotube plasmon resonators as a function of $L$ and doping level. (a) Attenuation as a function of $L$, with $t$ fixed to 165 nm and the nanotubes not intentionally doped. The attenuation is measured in a μ-FTIR apparatus with the light source polarized along the nanotube axis. From left to right, $L$ = 800 nm, 600 nm, 450 nm, 300 nm, 250 nm, and 200 nm. Inset: $Q$ factors as a function of $L$, reaching $Q = 9.0$ at $L = 200$ nm. The circles are color-coded to match the curves in the main panel. (b) Attenuation of a $t = 140$ nm, $L = 220$ nm film as a function of time after $NO_2$ exposure, which is a proxy for the doping level. The dashed lines represent line cuts plotted in (c). (c) Attenuation immediately after $NO_2$ exposure (maximum doping level), 1 day after $NO_2$ exposure (intermediate doping level). The solid blue line is the attenuation when the film has been annealed at 550 °C (minimal doping).

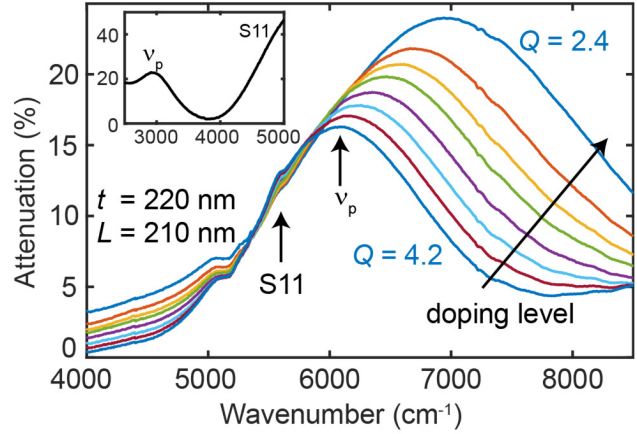

**Figure 3.** High-frequency (near-infrared) plasmon resonance of a thick ($t = 220$ nm) nanotube film at 8 different doping levels, measured in a UV-VIS-NIR spectrometer with an unpolarized light source. The curves represent time after exposure to $HNO_3$ vapor, with 10 minute intervals between curves, starting immediately after $HNO_3$ exposure. The $\nu_p$ peak can exceed the $S_{11}$ energy and be as high as 7000 cm$^{-1}$ (1.43 µm). Inset: Attenuation vs. wavenumber when the film is annealed (minimal doping).

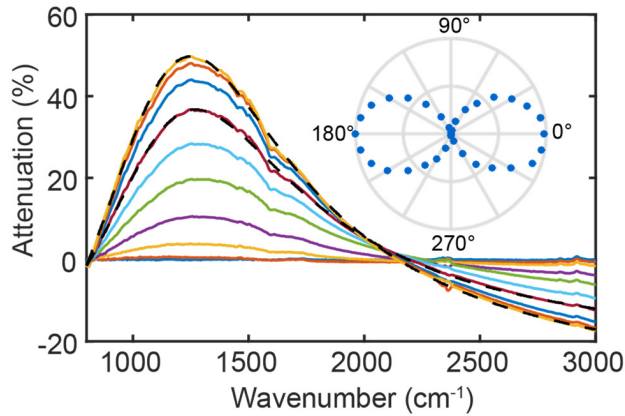

**Figure 4.** Plasmon resonance as a function of incident light polarization, with θ representing deviation from the nanotube alignment axis, taken at 10° intervals. A background curve representing is subtracted from all curves equally (see SI). The attenuation curves have excellent fits to Fano functions (black dashed lines, plotted for θ = 0° and θ = 30°). Inset: $I(\theta)$, fitted from the polarization-dependent attenuation using Eq. (1). The attenuation is fit to be 99% polarized along the axis of the nanotubes.

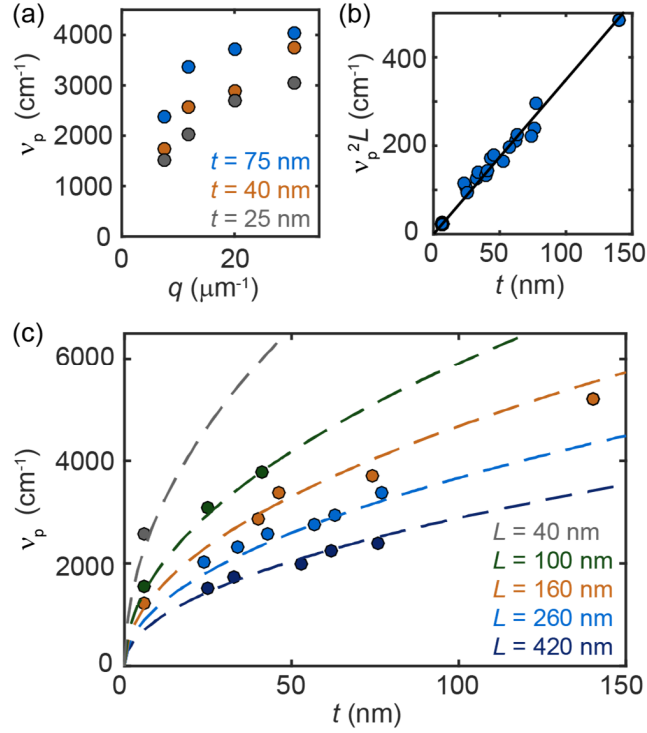

**Figure 5.** Evolution of $v_p$ with $q$ (= $\pi/L$) and $t$. (a) $v_p$ has a sublinear dependence on $q$. (b) When both $L$ and t are varied, $v_p^2 L$ has a linear relationship with $t$, indicating that $v_p$ has a good fit to Eq. 2. (c) $v_p$, plotted as a function of $t$. The fit (dashed lines) to Eq. 2 is a global fit to the data from all five curves, yielding $v_0 = 3335$ cm$^{-1}$. The three $t = 6$ nm data points are appropriated from Ref. 6. All the data in this figure are with highly doped nanotubes, measured immediately after NO$_2$ gas exposure.

**Supporting Information for**

**Strong and broadly tunable plasmon resonances in thick films of aligned carbon nanotubes**

Kuan-Chang Chiu[1,2], Abram L. Falk[1], Po-Hsun Ho[1], Damon B. Farmer[1], George Tulevski[1], Yi-Hsien Lee[2], Phaedon Avouris[1], Shu-Jen Han[1]

[1]IBM T. J. Watson Research Center, Yorktown Heights, New York, 10598, USA
[2]Dept. of Mat. Sci. and Eng., National Tsing-Hua University, Hsinchu, 30013 Taiwan

**Contents**

1. Detailed procedure for fabricating thick films of aligned carbon nanotubes.
2. Spectroscopy methods
3. Discussion of background subtraction and $Q$ factor calculation
4. Lower-frequency plasmon resonances in thinner films
5. Additional $L$-dependent series of plasmon resonances

**1. Detailed procedure for fabricating thick films of aligned carbon nanotubes.**

A vacuum filtration method was used to assemble the thick nanotube films[25]. For the experimental setup (Fig. 1(a) of main text), the key components are a 15 mL glass funnel and a fritted glass filter support with a silicone stopper (Millipore® XX1002500 glass microanalysis), both purchased from Fisher Scientific Company. The vacuum filtration method consists of four steps:

    1. Dispersing the carbon nanotubes in a surfactant solution.

    2. Vacuum-filtering the nanotube dispersion onto a filtration membrane.

    3. Transferring the CNT film onto a target substrate.

    4. Dissolving the membrane in a solvent.

An arc-discharge single-walled carbon nanotube powder with an average diameter of 1.4 nm and semiconducting purity of > 90% (P2-SWCNT, Carbon Solution) was dispersed in aqueous solution with sodium dodecylbenesulfonate (SDBS, Sigma-Aldrich) as a surfactant. The starting concentration was 0.4 mg/ml for nanotubes and 0.4% (wt/vol) for SDBS. The nanotube powder and surfactant was first dispersed by bath sonication for 15 minutes and then further sonicated with a tip sonicator for 45 min. Pulse sonication used a 3 sec/1 sec on/off duty cycle, and a circulating cooling water bath during to prevent heating of the dispersion. Next, the suspension was centrifuged for 1 hour at 38,000 r.p.m to remove large bundles from the suspension. The supernatant was extracted and then diluted to a final SDBS concentration of 0.08%.

The well-dispersed solution was filtered with polycarbonate filter membranes (Whatman Nuclepore track-etched polycarbonate hydrophilic membranes, 0.05 µm pore size) at a very low filtration speed of ~1 ml/hr. The filter membrane is treated with $O_2$ plasma for 30 seconds prior to deposition. The reduced filtration speed could be controlled by adjusting a needle valve and monitoring pressure with a differential vacuum pressure gauge.

The silicon substrates onto which we transferred the nanotube films consist of high-resistivity silicon with 10-nm of $HfO_2$, which was deposited by atomic layer deposition. For transferring the nanotube film, a droplet of water was placed on the target substrate. The as-deposited films on membranes were then attached on the target substrates with the side of nanotube film touching the substrate surface. A glass slide was used to cover the whole membrane with a small amount of pressure to make the film stick to the surface of the substrate firmly. Nitrogen gas was then used to dry the substrate.

To remove the membrane, the whole substrate was immersed into an organic solvent (N-methyl-2-pyrrolidone or chloroform) to dissolve the membranes. The samples were then annealed under a vacuum of $10^{-7}$ Torr at 500 degrees for thirty minutes to burn off any remaining organic residue.

## 2. Spectroscopy methods

The attenuation spectra were measured in a Bruker Nicolet 8700 µ-FTIR system. The diameter of the focused beam size is 25 µm, and each measurement was integrated for a 30-second period. Before each spectrum is taken, a background spectrum is taken, on an area of the chip in which the carbon nanotubes have been completely etched away. The spectrometer outputs absorbance relative to the background spectrum. We then calculate the attenuance ($A(\nu)$) as

$$A(\nu) = 1 - 10^{-absorbance(\nu)} \tag{S1}$$

To measure the high frequency plasmon resonances in thick films ($\nu_p > 6000$ cm$^{-1}$, see Fig. 3 of main text), which were out of the range of our FTIR system, we used a PerkinElmer Lambda 950 UV-VIS-NIR spectrometer. The films we prepared for use in this system were similar to the ones that we prepared for FTIR, except 1) Larger patterns (5 mm × 5 mm) were prepared, to accommodate the larger beam size of this spectrometer, and 2) glass substrates were used instead of silicon, to reduce the substrate absorption in this higher energy regime.

## 3. Background subtraction and *Q* factor calculation

As mentioned above, all of the measurements in this paper were taken relative to a background spectrum, where the background consisted of a region of the substrate with nanotubes that have been etched away. Nonetheless, in some cases, features that we do not attribute to the plasmon resonance of the etched nanotubes remain in our spectra. For instance, $S_{11}$ absorption can clearly be seen in Fig. 2(c) and in Fig. 3.

When calculating *Q* factors in this manuscript, we always calculate full-width half-max (FWHM) relative to only the shape of the plasmon resonance ($\nu_p$). That is, we subtract off any spectral features that are not part of $\nu_p$, and calculate the FWHM relative to the base of only the $\nu_p$ peak. Likewise, when quoting peak attenuation value, we quote the peak height of $\nu_p$ after subtracting off any non-$\nu_p$ attenuation features.

For the spectra plotted in Fig. 2(a), a constant background is subtracted from each curve to make the minimum attenuation line up with 0%. This procedure eliminates any potential contribution from $S_{11}$ attenuation and unetched nanotubes to the peak attenuation. The spectra plotted in Figs. 2(b-c) and 3 are likewise plotted as measured, except for the subtraction of a small (2%-5%) constant attenuation.

In Fig. 4, the polarization sensitive data reveal the rising edge of a low energy peak (Fig. S1(a)). This rising edge likely corresponds to the plasmon resonance of unetched nanotubes, possibly etch-resistant nanotube bundles. We fit these data to a sum of two Fano functions, with the model that the higher energy one would correspond to the etched nanotubes ($\nu_p$) and the lower energy one would correspond to the unetched nanotubes. We find this lower energy rising edge to be well fit to the Fano function (Eq. 1), with parameters $\nu_0 = 960$ cm$^{-1}$, $F = 0.4$, and $\gamma = 3000$ cm$^{-1}$ (dashed black line in Fig. S1(a)). After subtracting this curve from all of the polarization-dependent data (Fig. S1(b)), we then refit the data to a single Fano function, with excellent fitting results (Fig. 4 in main text).

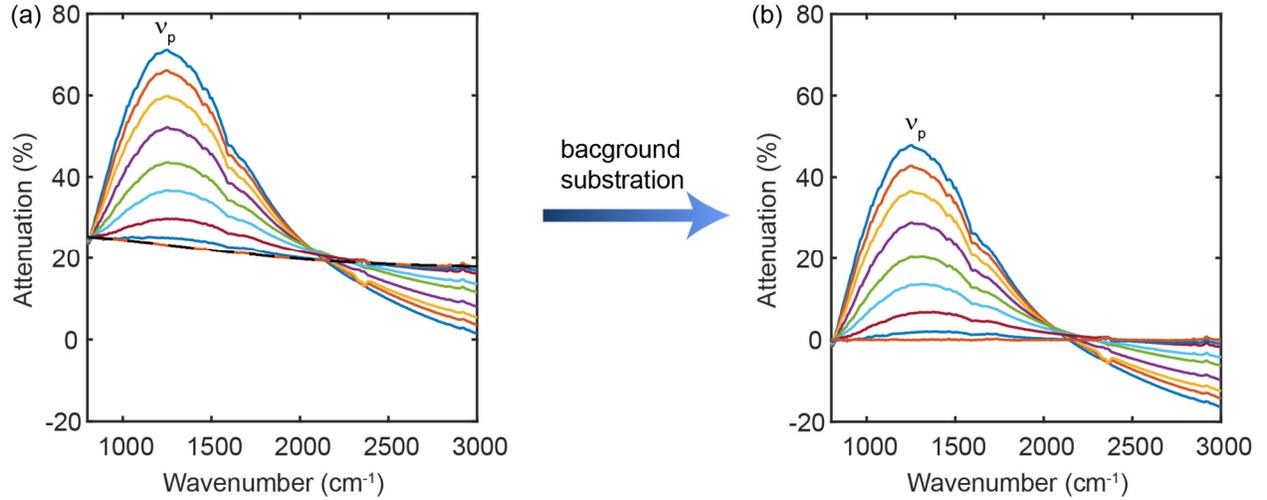

**Figure S1.** Background subtraction for polarization dependent data. (a) Attenuations as a function of incident light polarization, with θ representing deviation from the nanotube alignment axis, taken at 10° intervals. The dashed black curve is the fit to Eq. 1 of the rising edge of the low energy feature (fit to a peak wavenumber of 960 cm$^{-1}$). (b) Polarization-dependent attenuation after subtracting off the background curve. These are the curves displayed in Fig. 4 of the main text.

## 4. Lower-frequency plasmon resonances in thinner films

As discussed in the main text, $\nu_p$ is lower for thinner films. In a relatively thin $t = 40$ nm film, $\nu_p$ ranged from approximately 1000 cm$^{-1}$ up to 3800 cm$^{-1}$ (Fig. S2). When $\nu_p > 2000$ cm$^{-1}$, the lineshapes are approximately symmetrical. However, when $\nu_p < 2000$ cm$^{-1}$, the lineshapes become asymmetrical, and a sharp secondary peak at 760 cm$^{-1}$ strengthens (marked "LO" in Fig. S2).

In previous work that studied plasmons in thin nanotube films[6], a very similar phenomenon was observed, and the asymmetric lineshapes and secondary peak were understood to result from plasmon-coupled longitudinal phonons in the SiO2 substrate. In this work, we have a $HfO_2$ substrate, not an $SiO_2$ one, but the lineshapes can be understood as deriving from an analogous plasmon-coupled longitudinal optical (LO) phonon in the $HfO_2$ substrate.

When measured immediately after heavy p-type doping through $HNO_3$ vapor, the $\nu_p$ resonances are approximately 1.5× higher than comparable resonances in the nominally undoped nanotube film. Interestingly, this 1.5× factor is approximately the same as the factor observed in thin ($t = 6$ nm) nanotube films[6] and significantly less than the 2.3× factor we observe for the thick nanotube films (Fig. 3 in main text). It suggests that inter-nanotube coupling is a function of doping level, thereby shifting $\nu_p$ in a thick film even more than $\nu_p$ in an isolated nanotube or thinner film would shift under a similar doping-level differential.

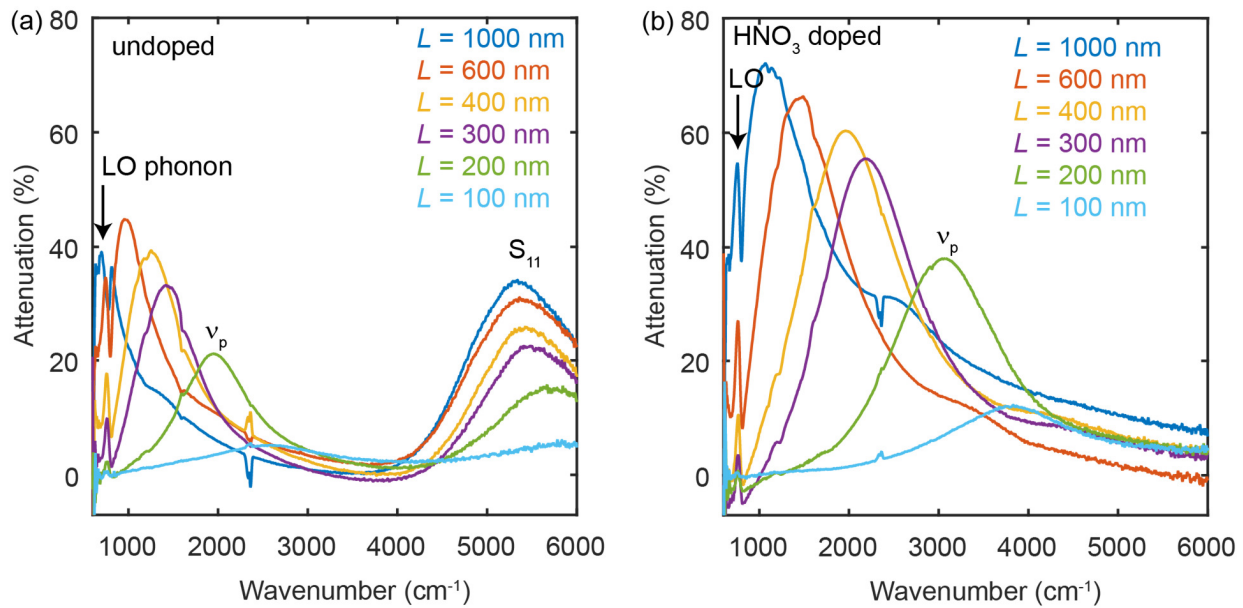

**Figure S2.** Lower frequency plasmon resonances on a thinner ($t = 40$ nm) carbon-nanotube film. (a) Attenuation spectra as a function of $L$ when the nanotubes are undoped. (b) Attenuation spectra as a function of $L$ when the films are doped with $HNO_3$ vapor.

## 5. Additional $L$-dependent series of plasmon resonances

As a complement to Fig 2(a), Fig. S3 presents a similar series of length-dependent plasmon resonances.

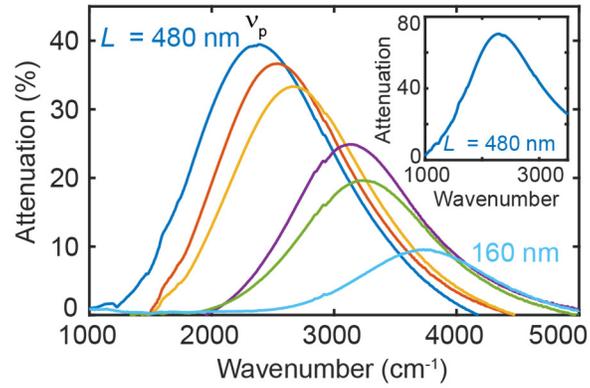

**Figure S3.** Attenuation as a function of $L$, with $t$ fixed to 70 nm and the nanotubes in a highly doped state immediately following $NO_2$ exposure. The attenuation is measured in a µ-FTIR apparatus with an unpolarized light source. From left to right, $L$ = 480 nm, 410 nm, 350 nm, 260 nm, 230 nm, and 160 nm, with corresponding $Q$ factors of 1.8, 2.0, 2.1, 2.6, 2.6, and 3.0. Inset: Attenuation when incident light is polarized along the nanotube axis, reaching 70% peak attenuation.